\providecommand{\columnwiselinenumbersfalse}{}
\providecommand{\runningpagewiselinenumbers}{}
\providecommand{\linenumbers}{}

\documentclass[letter,desactivate]{aa}

\usepackage[varg]{txfonts}
\usepackage{graphicx}
\usepackage{natbib}
\usepackage{xcolor}
\usepackage[normalem]{ulem}
\bibpunct{(}{)}{;}{a}{}{,}

\begin{document}

\title{Polar-structure galaxies beyond the local Universe: \\first results from Euclid}

\author{J. A. Guerrette\inst{1}\corrauth{vinnieguerrette@gmail.com} 
    \and A. V. Mosenkov\inst{1}\email{mosav@byu.edu}
    \and S. K. H. Bahr\inst{1}\email{hsenecak@byu.edu}
    \and I. H. Jensen\inst{1}\email{ihj22@byu.edu}
    \and J. X. George\inst{2}\email{jonahg17@terpmail.umd.edu}
    \and T. E. Spigarelli\inst{1}\email{theaes@byu.edu}
    \and R. P. Smith\inst{3}\email{rpsmith4@illinois.edu}
    \and B. T. Burton\inst{1}\email{burton77@byu.edu}
    \and K. W. Beckstead\inst{1}\email{kevanwb@byu.edu}
    \and J. D. Seguine\inst{1}\email{jseguine@byu.edu}
    }

\institute{Department of Physics and Astronomy, N283 ESC, Brigham Young University, Provo, UT 84602, USA
  \and Department of Astronomy, 4296 Stadium Drive, Astronomy Dept Room 1113, University of Maryland, College Park, MD 20742, USA
  \and Department of Physics, 1110 West Green Street, University of Illinois at Urbana-Champaign, Urbana, IL 61801, USA}

\date{Received date / Accepted date}

\abstract{
Galaxies hosting strongly misaligned stellar and gas structures offer a direct window into external accretion and the buildup of angular momentum through multiple spin components. Such polar-structure systems are intrinsically rare and, until now, have been identified almost entirely at low redshift. The Euclid Quick Data Release (Q1), with its uniform, space-based VIS imaging, provides an unprecedented opportunity to locate these objects over wide areas and at much greater distances. We present the first visual search for polar-structure galaxy (PSG) candidates in Euclid VIS data, spanning approximately 10\% of the Q1 survey footprint. This search yields the largest high-redshift PSG candidate sample assembled to date, including roughly 100 high-confidence systems with over 50 at $z \gtrsim 0.3$, along with additional populations of face-on rings and polar tidal structures (PTS). Robust candidates are detected out to redshifts of $z\sim1$, demonstrating that polar structures can be identified well beyond the local Universe. Although the present sample is not yet sufficient for quantitative occurrence rate or evolutionary measurements, it enables initial classification and basic structural characterization of PSGs at intermediate redshift. These results lay the groundwork for future statistical studies with forthcoming Euclid data releases and illustrate Euclid’s power for rare-object science across cosmic time.
}



\maketitle

\section{Introduction} 
\label{sec:intro}

Polar-structure galaxies (PSGs) are galaxies that host large-scale components significantly misaligned with respect to the primary stellar body. These structures appear in a variety of morphologies, including stellar and gaseous polar rings \citep{Whitmore1990, Moiseev2011}, polar bulges \citep{Corsini2012, Reshetnikov2015}, polar halos \citep{Mosenkov2020, Mosenkov2022}, and asymmetric polar tidal structures \citep{Mosenkov2024, Bahr2026}.

Polar structures are widely interpreted as signatures of external accretion or interaction, with no known secular formation methods to form large-scale, long-lived polar structures \citep{Reshetnikov1997, Bournaud2003}. Minor and major mergers \citep{Ordenes2016, 
Quiroga2020}, tidal accretion from companion systems \citep{Martinez2010, Khalid2024}, and gas infall from cosmological filaments \citep{Brook2008, Danovich2015} can all deposit material with angular momentum misaligned from that of the host galaxy.

Recent deep, wide-area optical imaging has greatly expanded the known population of PSGs in the nearby Universe. The Catalog Of Unusual Galaxies with polar Structures \citep[COUGS-DESI;][]{Bahr2026}, based on the DESI Legacy Imaging Surveys \citep{Dey2019}, provides the largest and most homogeneous census to date, identifying nearly 3{,}000 candidates and increasing the known sample by almost an order of magnitude. This work demonstrated that polar structures are more common and morphologically diverse than previously recognized \citep{Whitmore1990, Moiseev2011}. However, COUGS remains largely confined to the local Universe ($z \lesssim 0.3$) and shows a significant selection bias beyond $z \gtrsim 0.05$, where polar structures become increasingly difficult to identify reliably.

This limitation becomes even more severe at higher redshift. Beyond $z \sim 0.3$, PSGs remain poorly characterized because decreasing angular size and cosmological surface-brightness dimming \citep{Whitney2020} make polar structures progressively harder to detect and classify. As a result, only one PSG candidate is currently known at $z \sim 1$ \citep{Reshetnikov2007}, and the evolution of the PSG population beyond the local Universe remains essentially unexplored.

The Euclid mission provides the spatial resolution and depth needed to overcome these limitations. With $0.18\arcsec$ $I_E$-band resolution \citep{Euclid2025VIS} and an average photometric depth approaching $30~\mathrm{mag~arcsec^{-2}}$ (see Sect.~\ref{sec:sample}), Euclid enables the identification of faint, small-angular-scale polar structures at substantially higher redshift, extending visually-selected PSG samples into the intermediate-redshift Universe with uniform space-based imaging.

Expanding the PSG census to higher redshift is key to probing the role of accretion and angular-momentum misalignment in galaxy evolution. In this Letter, we present the first sample of visually identified PSG candidates and ring galaxies from randomly selected VIS tiles comprising $\sim10$\% of Euclid's first Quick Data Release (Q1; \citealt{Euclid2025Q1}) and demonstrate Euclid's ability to resolve polar structures and rings out to $z \sim 1$, thereby laying the groundwork for future statistical studies of the PSG population.

Throughout this article we adopt a standard flat $\Lambda$CDM cosmology with $\Omega_m=0.3$, $\Omega_{\Lambda}=0.7$, and $H_0=70$~km\,s$^{-1}$\,Mpc$^{-1}$.


\section{Data and Sample Selection} 
\label{sec:sample}

Euclid’s Q1 covers a total area of $63.1 \deg^2$, distributed across three deep fields: Euclid North, Fornax, and Euclid South. An additional $0.5\deg^2$ was observed toward a ``Dark Cloud'' region in the Orion Nebula and used for calibration, though that was omitted from our analysis. Imaging was obtained with the VIS broad-band optical imager and the Near-Infrared Spectrometer and Photometer (NISP). VIS observes in a single broad optical band ($I_E$), while NISP provides imaging in the $Y_E$, $J_E$, and $H_E$ bands. The data are delivered as individual tiles that together form mosaics across each field.

From the full Q1 footprint, we randomly selected 26 VIS MER mosaic tiles from the ESA Euclid Science Archive, corresponding to approximately $11\%$ of the total surveyed area ($6.86\,\deg^2$). In this work, we restrict our analysis to the VIS imaging. The optical band is particularly well suited for identifying stellar polar structures, which are often faint yet morphologically distinct. The VIS data provide a pixel scale of $0.1\arcsec\,\mathrm{pix}^{-1}$, significantly finer than the $0.3\arcsec\,\mathrm{pix}^{-1}$ sampling of the NISP near-infrared wavebands. Moreover, the VIS point-spread function has a full width at half maximum (FWHM) of $\sim0.18\arcsec$, compared to values of up to $\sim0.48\arcsec$ in the NISP bands \citep{Euclid2025FWHM}. Based on our measurements, the average photometric depth across the selected tiles, calculated using 1000 random $10\arcsec \times 10\arcsec$ boxes per tile, is $29.9~\mathrm{mag\,arcsec^{-2}}$ at $3\sigma$. These characteristics enable the detection and robust morphological characterization of small-angular-scale, misaligned stellar structures at intermediate redshift.

PSGs are intrinsically rare and morphologically diverse, making automated identification challenging, particularly in the absence of large, well-defined training samples, especially at intermediate redshifts. For this reason, we adopted a fully visual identification strategy.

To ensure a systematic search, each VIS tile was divided into $3.2\arcmin \times 1.7\arcmin$ subtiles, with the exact number of subtiles depending on the angular size of the parent tile. These subtiles were then reviewed sequentially by multiple classifiers so that each tile received at least two independent inspections. Galaxies exhibiting large-scale polar components or rings of any kind were flagged for further inspection. The selected candidates were re-examined by a different classifier to confirm their morphological classification. In our selection of polar structures, we do not impose a strict requirement of exact orthogonality; instead, we include objects that exhibit a substantial tilt between two stellar structures ($\gtrsim40^\circ$; \citealt{Smirnova2013}). The spatial distribution of the candidates within the selected Q1 tiles is shown in Fig.~\ref{fig:skymaps}.

\begin{figure}
    \centering
    \includegraphics[width=0.48\linewidth]{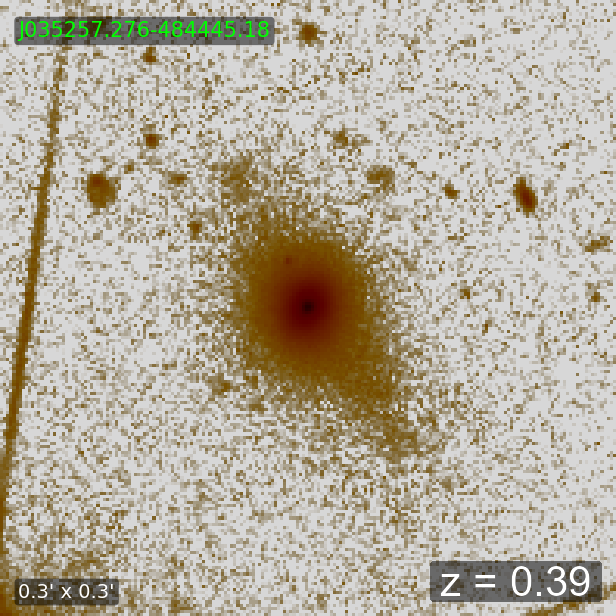}
    \includegraphics[width=0.48\linewidth]{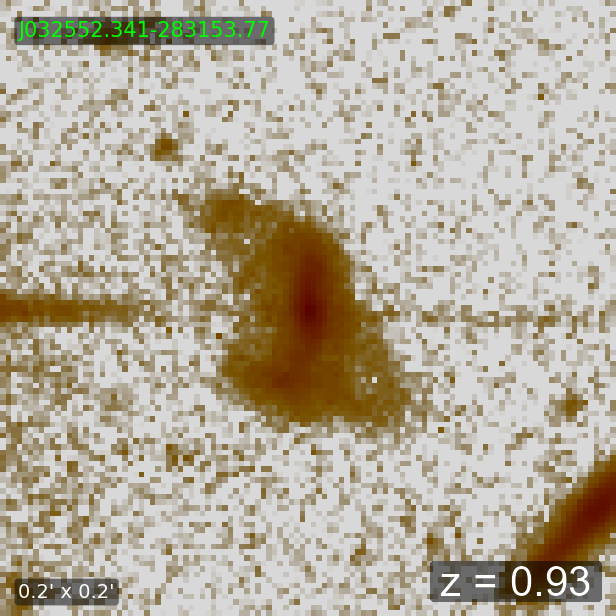}
    \\
    \includegraphics[width=0.48\linewidth]{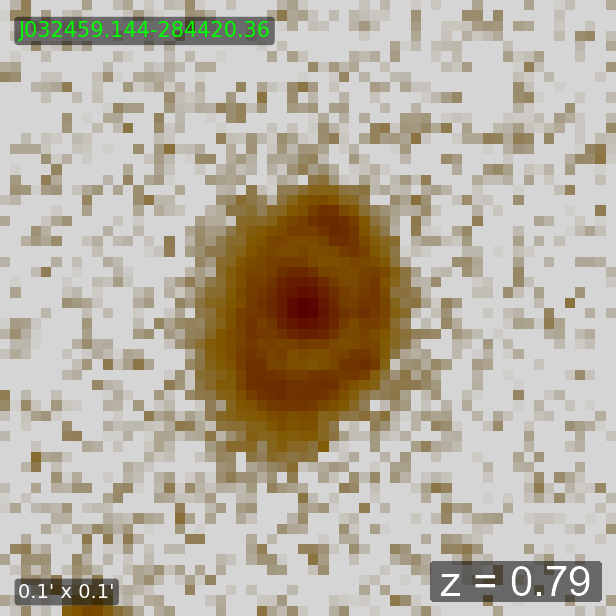}
    \includegraphics[width=0.48\linewidth]{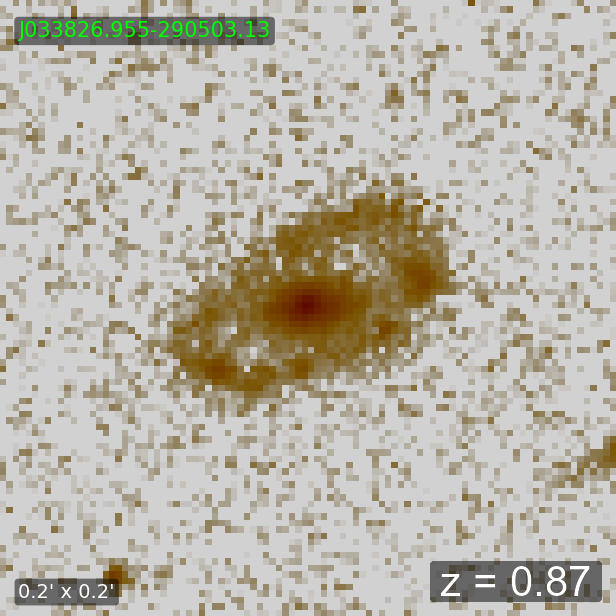}
    \\
    \includegraphics[width=0.48\linewidth]{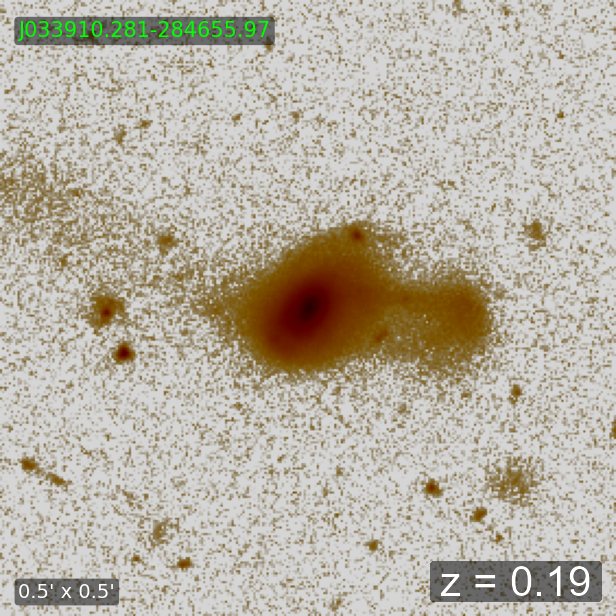}
    \includegraphics[width=0.48\linewidth]{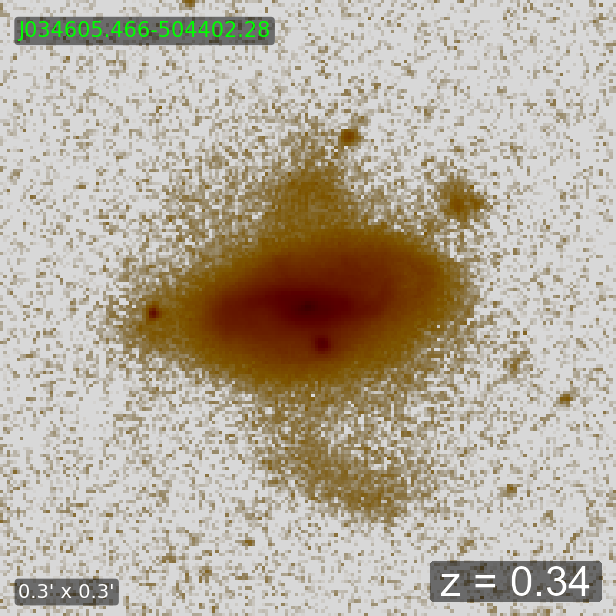}
    \caption{Representative Euclid VIS cutouts with the outer structures enhanced for visibility (see Appendix~\ref{sec:Atlas}) illustrating examples from the three morphological categories identified in this work: high-confidence PSG candidates (top row), face-on ring galaxies (middle row), and polar tidal structure galaxies (bottom row). The labels on each image represent the J2000 coordinates (top left), the angular size of the cutout (in arcminutes; bottom left), and the photometric redshift (bottom right).}
    \label{fig:sample_gals}
\end{figure}

The final sample comprises three main morphological groups: (1) high-confidence (``best'') PSGs; (2) ringed galaxies, typically observed face-on; and (3) polar tidal structures (PTS), characterized by irregular, asymmetric, or disturbed polar-structure morphologies indicative of ongoing or recent accretion or minor mergers, some of which may lead to quasi-stable PSGs. Ring-like galaxies were retained as a separate, lower-confidence category because some face-on polar structures can appear similar to more conventional barred ring galaxies in imaging data alone. Since ring morphologies may also arise from collisional or resonance-related processes, limited spatial resolution and projection effects often prevent a secure distinction between these scenarios. Even if some of the selected ring galaxies are not true PSGs, some ring galaxies can form through similar evolutionary processes (such as mergers and accretion) and therefore offer complementary insight when studied alongside PSGs. We emphasize that this classification is inherently subjective and that our sample comprises candidate PSGs whose polar nature has not yet been confirmed kinematically.

\begin{figure*}[htbp]
    \centering
    \includegraphics[width=0.42\linewidth]{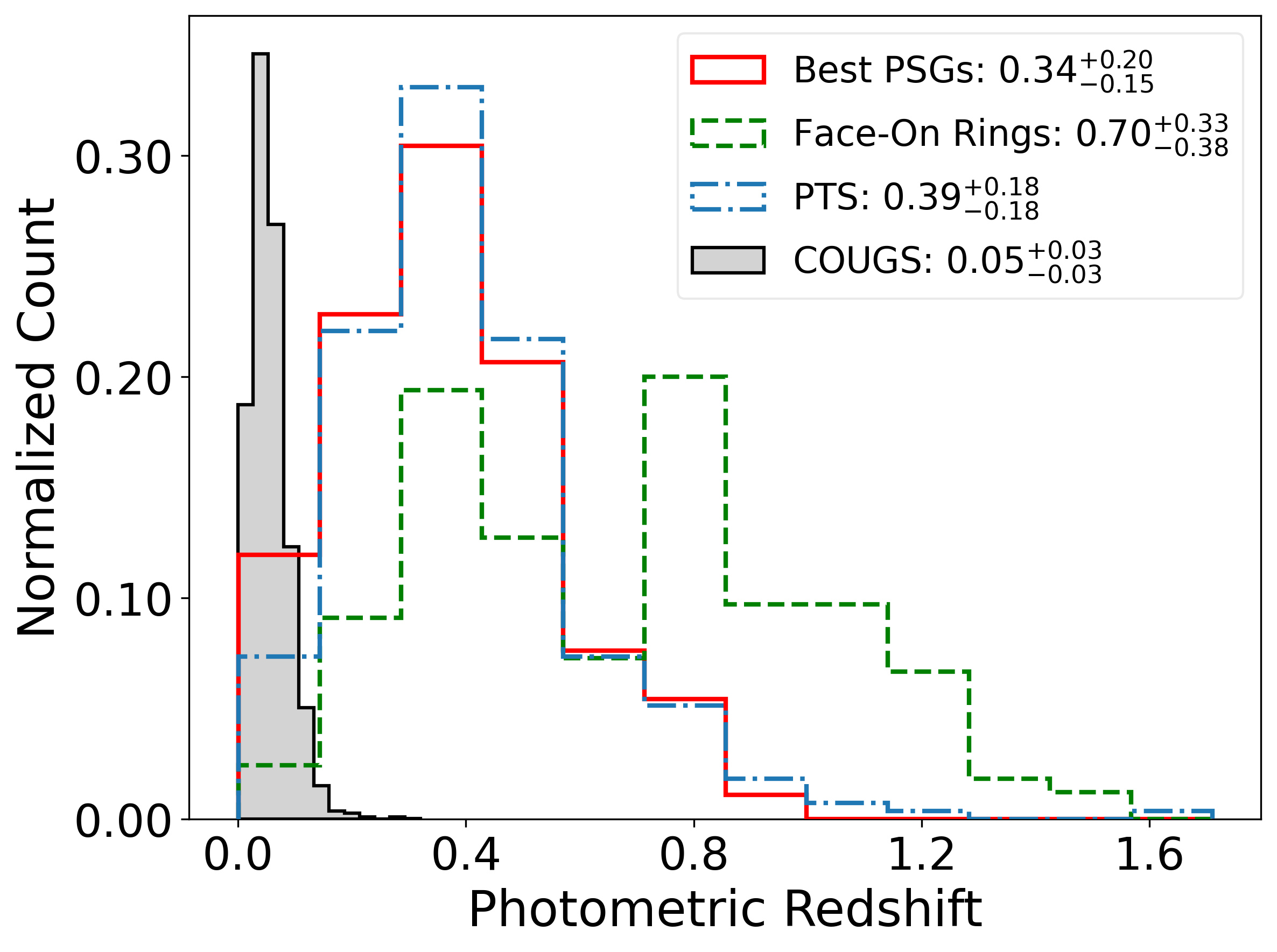}
    \includegraphics[width=0.42\linewidth]{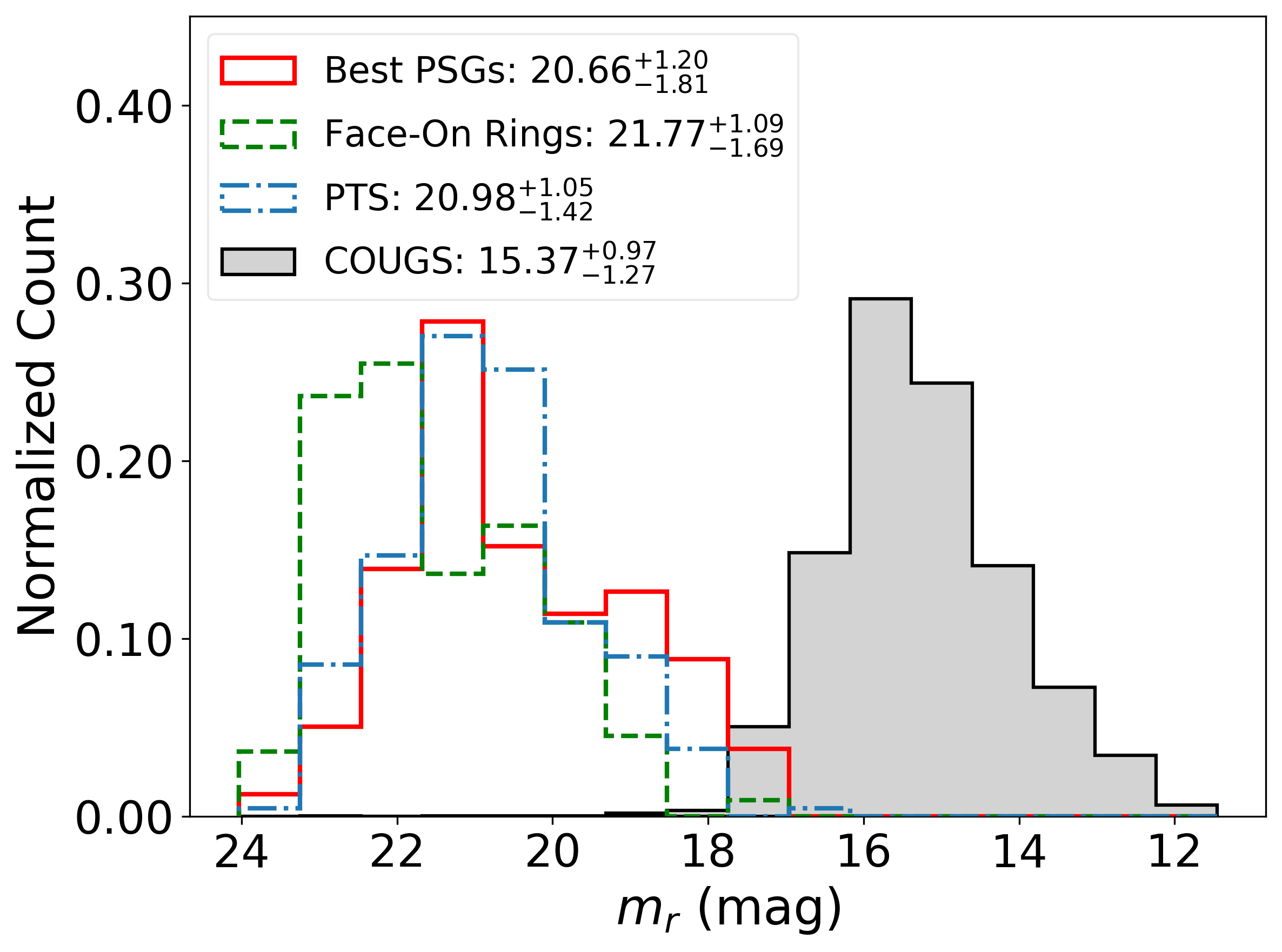}
    \includegraphics[width=0.42\linewidth]{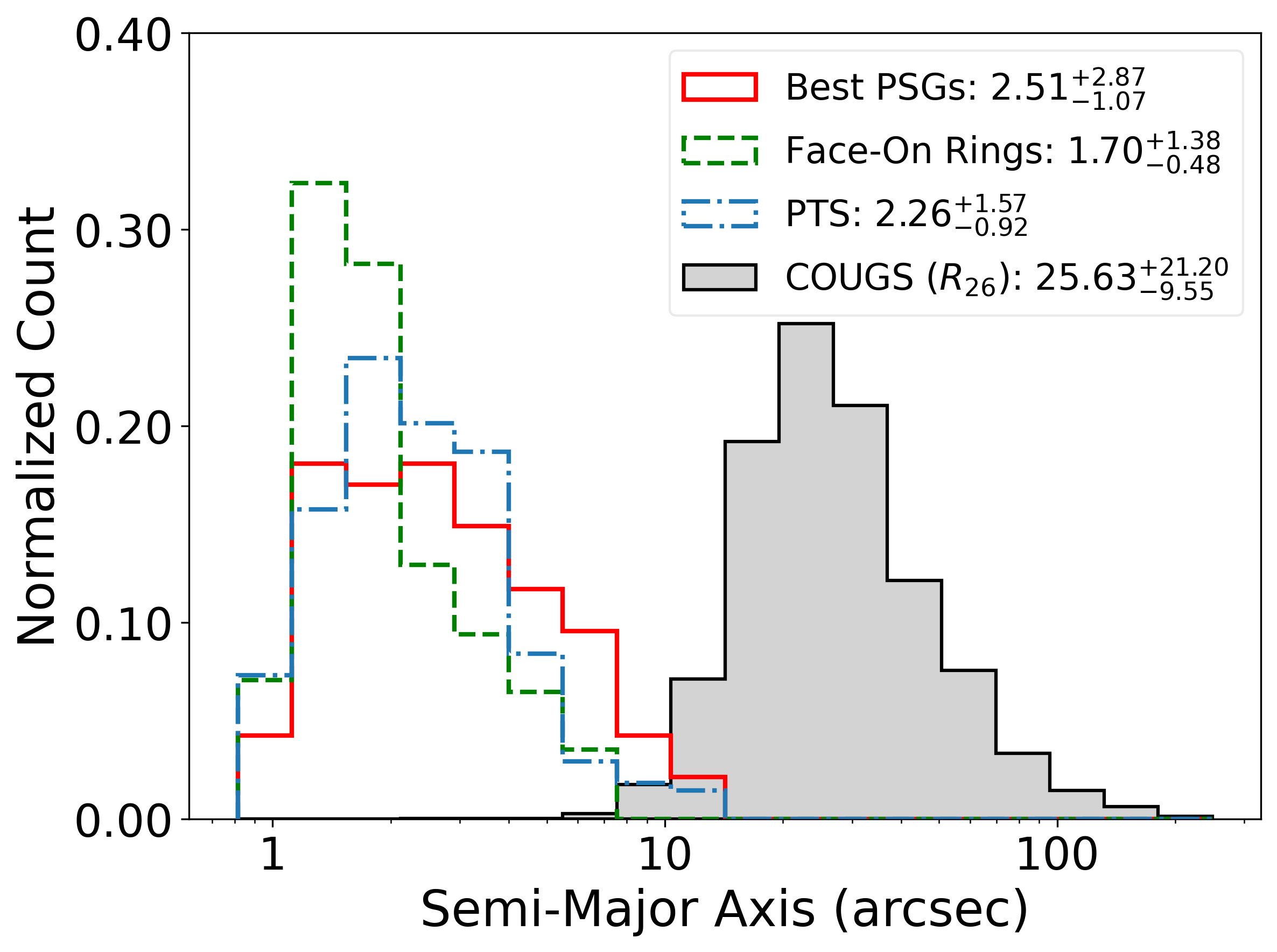}
    \includegraphics[width=0.42\linewidth]{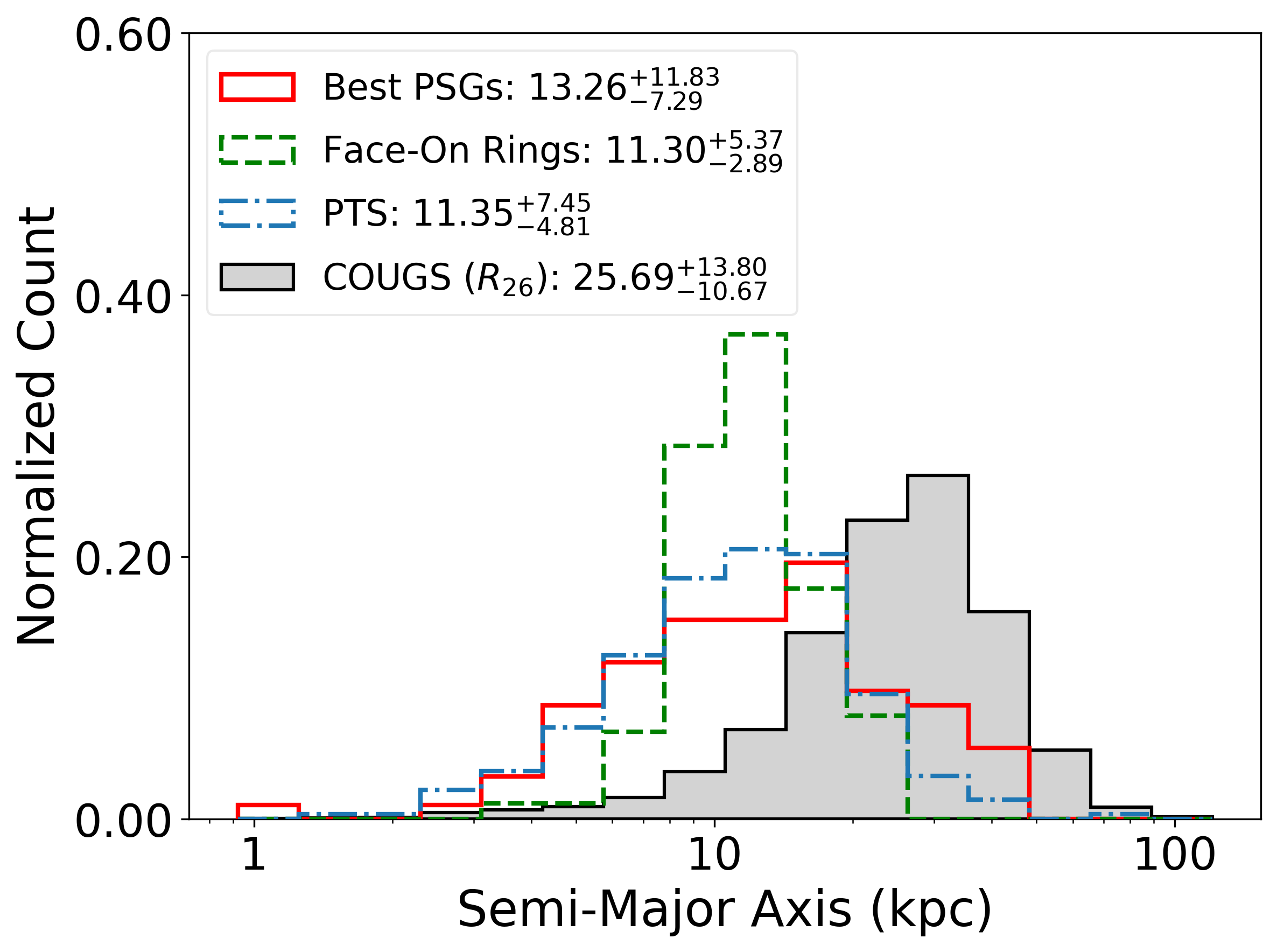}
    \caption{Normalized count histograms displaying the photometric redshift (obtained from the Euclid PHZ tables), \textit{r}-band apparent magnitude (obtained from the Euclid MER tables), and size distributions (semi-major axis of an ellipse fit to the 26th isophote, $R_{26}$; both angular and physical) for our three samples. The shaded bars in each plot represent the parameter distribution from the COUGS-DESI catalog for comparison \citep{Bahr2026}. The plot legends show the median value of each sample, along with the 16th and 84th percentiles.}
    \label{fig:histograms}
\end{figure*}

Representative VIS cutouts illustrating these categories are presented in Fig.~\ref{fig:sample_gals}. Table 1, available electronically, contains the properties of the PSGs analyzed in this work. Column 1 gives the morphological sample classification, Column 2 gives the polar structure classification, and Column 3 lists the internal galaxy identifier. Column 4 gives the Euclid Q1 tile identifier. Columns 5 and 6 provide the visually measured J2000 right ascension and declination of the galaxy center. Columns 7-10 give the adopted semi-major and semi-minor axes in arcseconds and kiloparsecs, while Column 11 gives the angular scale used for the physical-size conversion and Column 12 the adopted position angle. Columns 13-15 provide the matched Euclid source identifier and its J2000 coordinates. Columns 16-19 give the Euclid MER \textit{r}-band template-fit extended-source flux, the corresponding apparent \textit{r}-band magnitude, the MER total detection-band flux, and the Euclid VIS Kron apparent magnitude, respectively. Column 20 gives the median photometric redshift from the Euclid PHZ catalog.

The inspection yielded 94 high-confidence PSG candidates. In addition, 170 face-on ring systems and 273 polar tidal structure candidates were identified. In this Letter, we focus primarily on the high-confidence PSGs, which we refer to as the Best sample. The auxiliary samples are included to capture the broader diversity of ringed and misaligned morphologies present in the dataset. They serve as a comparative reference for the Best polar-structure candidates and provide a foundation for future investigations of related structural configurations.


\section{Results and Discussion} 
\label{sec:results}

Fig.~\ref{fig:histograms} shows the distributions of photometric redshift, apparent \textit{r}-band magnitude, angular size, and physical size for our three Euclid PSG samples, together with the COUGS-DESI catalog for comparison \citep{Bahr2026}. These comparisons should be interpreted cautiously because the higher redshifts of the Euclid galaxies shift the observed \textit{r}-band and size measurements to bluer rest-frame wavelengths than those sampled by COUGS-DESI.

Photometric redshifts from the Euclid PHZ table \citep{Euclid2025PHZ, https://doi.org/10.26131/irsa604} indicate that all three subsamples extend well beyond the local Universe, with many candidates at $z_{\rm phot} \gtrsim 0.5$ and the highest-confidence PSGs reaching $z_{\rm phot}=0.93$. This demonstrates that Euclid VIS imaging can resolve visually identifiable polar structures out to intermediate redshift, substantially beyond the regime previously accessible to ground-based surveys.

The apparent \textit{r}-band magnitudes, derived from the \texttt{flux\_r\_ext\_decam\_templfit} measurements in the Euclid MER table \citep{Euclid2025MER, https://doi.org/10.26131/irsa602}, show that PSG candidates can be identified over a broad brightness range extending significantly fainter than the COUGS-DESI sample. This suggests that strongly misaligned stellar structures remain detectable even in relatively faint systems.

Semi-major axis sizes were measured from elliptical fits to a fixed isophote of $\mu_\mathrm{VIS}=26\,\mathrm{mag\,arcsec^{-2}}$, rather than from Kron radii, in order to better trace extended low-surface-brightness polar structures. The Best and PTS samples have comparable angular-size distributions, with median semi-major axes of $2.51\arcsec$ and $2.26\arcsec$, respectively, while the ring sample has a smaller median value of $1.70\arcsec$, likely because it extends to higher redshift. By comparison, the COUGS-DESI galaxies are approximately an order of magnitude larger in angular size due to their much lower redshifts. Since the VIS PSF FWHM is only $0.18\arcsec$, even the smallest systems in our sample are resolved over several PSF elements, enabling reliable identification of genuinely misaligned structures.

The physical-size distributions of the three Euclid subsamples are broadly similar, although the Best PSGs span the widest range of sizes. COUGS-DESI galaxies tend to be physically larger; however, direct comparison is complicated by rest-frame wavelength differences and cosmological surface-brightness dimming. For $z_{\rm med} = 0.34$, the expected dimming in the AB system is $\Delta \mu = 2.5\log_{10}\!\left((1 + z_{\rm med})^3 \right) \approx 0.95~\mathrm{mag~arcsec^{-2}}$

The broad redshift range probed by our sample also has important implications for PSG formation and evolution. Since PSGs are generally associated with external accretion and mergers, their occurrence may be linked to the increasing merger rate \citep{Lotz2011, Rodriguez2015} and higher gas fractions in galaxies at earlier cosmic epochs \citep{Tacconi2018}. At the same time, PSGs in quasi-equilibrium appear to preferentially inhabit relatively low-density environments \citep{Finkelman2012, Savchenko2017}, likely because high galaxy velocities in clusters suppress mergers and repeated interactions can disrupt fragile polar structures \citep{Savchenko2017}. JWST observations have shown that rotating disk galaxies were already in place by $z\gtrsim5$ \citep[see e.g.][]{Nelson2024}. Such systems could support the formation of polar structures through continued accretion and misaligned gas inflow \citep{Brook2008}. The PSG occurrence rate may therefore evolve with redshift, although the present sample is neither large nor statistically complete enough to test this possibility robustly.

These results demonstrate the potential of Euclid for studying multi-spin galaxy assembly across cosmic time. Since this Letter examines only $\sim11\%$ of the Euclid Q1 data, extending the analysis to the full Quick Data Release and future Euclid releases could yield thousands of PSG candidates over a broad redshift range. The visually identified sample presented here will also provide an expanded training set for convolutional neural networks designed for PSG identification \citep{2025A&A...702A.258D, Bahr2026}, improving automated searches at higher redshift.

Follow-up PSG studies could also be carried out with facilities such as JWST, the Nancy Grace Roman Space Telescope, and the Extremely Large Telescope, whose sensitivity and spatial resolution will enable studies of faint structures in the early Universe. However, the galaxies presented in this Letter should presently be regarded as PSG candidates rather than confirmed systems, since imaging alone cannot unambiguously establish the physical nature of the misaligned structures. Kinematic confirmation is therefore required \citep{Whitmore1990, Moiseev2011}, motivating future spectroscopic observations with instruments such as VLT/MUSE and Keck/MOSFIRE.

\section{Conclusions} 
\label{sec:conclusions}

Euclid's first quick data release revealed a sample of candidates for PSGs at low to intermediate redshifts. This marks the discovery of the first large sample of PSGs out to $z \sim 1$. These galaxies span the whole redshift range as well as a wide range of apparent magnitudes, highlighting the ability of Euclid's photometrically deep imaging to identify PSGs to cosmically large distances.

The presence of $\sim100$ high-confidence PSG candidates within $<7\deg^2$ of Euclid Q1 across a range of redshifts implies that polar structures have been present throughout a significant portion of the history of the Universe. This shows that these galaxies, though rare, are not restricted to the local Universe. Instead, their presence out to intermediate redshifts tells us that polar structures were already present at that epoch, though exactly how common they were will be determined by future occurrence rate estimates.

Observing PSGs at these early times opens a new window into galaxy formation, accretion, and angular momentum alignment. Since polar structures almost guarantee external influences on the host galaxy, studying them earlier in the history of the Universe will allow us to constrain the formation and evolution mechanisms necessary to form highly tilted angular momentum components in these early times.

\begin{acknowledgements}

This work has made use of the Quick Release (Q1) data from the Euclid mission of the European Space Agency (ESA). The VIS imaging used in this work was obtained from the ESA Euclid Science Archive. This research has also made use of the NASA/IPAC Infrared Science Archive, which is funded by the National Aeronautics and Space Administration and operated by the California Institute of Technology. We acknowledge the use of Euclid Q1 data and credit ESA Euclid/Euclid Consortium/NASA/Q1-2025.

The authors acknowledge the use of ChatGPT \citep{ChatGPT2026}, developed by OpenAI, for assistance with manuscript editing, revision, and clarity. All AI-assisted text was reviewed, edited, and verified by the authors, who take full responsibility for the content of this manuscript.
\end{acknowledgements}

\bibliographystyle{aa}
\bibliography{main}

@ARTICLE{Bahr2026,
       author = {{Bahr}, Seneca K.~H. and {Mosenkov}, Aleksandr V. and {Guerrette}, Jacob A. and {Jensen}, Isaac H. and {George}, Jonah X. and {Spigarelli}, Thea E. and {Smith}, Ryan P. and {Burton}, Brandon T. and {Beckstead}, Kevan W. and {Seguine}, Jonah D. and {Casper}, Harrison K.},
        title = "{COUGS-DESI: A Catalog of Unusual Galaxies with Polar Structures in the DESI Legacy Imaging Surveys}",
      journal = {arXiv e-prints},
         year = 2026,
        month = jan,
          eid = {arXiv:2601.02579},
        pages = {arXiv:2601.02579},
          doi = {10.48550/arXiv.2601.02579},
archivePrefix = {arXiv},
       eprint = {2601.02579},
 primaryClass = {astro-ph.GA},
       adsurl = {https://ui.adsabs.harvard.edu/abs/2026arXiv260102579B}
}

@ARTICLE{Lotz2011,
       author = {{Lotz}, Jennifer M. and {Jonsson}, Patrik and {Cox}, T.~J. and {Croton}, Darren and {Primack}, Joel R. and {Somerville}, Rachel S. and {Stewart}, Kyle},
        title = "{The Major and Minor Galaxy Merger Rates at z < 1.5}",
      journal = {\apj},
         year = 2011,
        month = dec,
       volume = {742},
       number = {2},
          eid = {103},
        pages = {103},
          doi = {10.1088/0004-637X/742/2/103},
archivePrefix = {arXiv},
       eprint = {1108.2508},
 primaryClass = {astro-ph.CO},
       adsurl = {https://ui.adsabs.harvard.edu/abs/2011ApJ...742..103L}
}

@ARTICLE{Rodriguez2015,
       author = {{Rodriguez-Gomez}, Vicente and {Genel}, Shy and {Vogelsberger}, Mark and {Sijacki}, Debora and {Pillepich}, Annalisa and {Sales}, Laura V. and {Torrey}, Paul and {Snyder}, Greg and {Nelson}, Dylan and {Springel}, Volker and {Ma}, Chung-Pei and {Hernquist}, Lars},
        title = "{The merger rate of galaxies in the Illustris simulation: a comparison with observations and semi-empirical models}",
      journal = {\mnras},
         year = 2015,
        month = may,
       volume = {449},
       number = {1},
        pages = {49-64},
          doi = {10.1093/mnras/stv264},
archivePrefix = {arXiv},
       eprint = {1502.01339},
 primaryClass = {astro-ph.GA},
       adsurl = {https://ui.adsabs.harvard.edu/abs/2015MNRAS.449...49R}
}

@ARTICLE{Tacconi2018,
       author = {{Tacconi}, L.~J. and {Genzel}, R. and {Saintonge}, A. and {Combes}, F. and {Garc{\'\i}a-Burillo}, S. and {Neri}, R. and {Bolatto}, A. and {Contini}, T. and {F{\"o}rster Schreiber}, N.~M. and {Lilly}, S. and {Lutz}, D. and {Wuyts}, S. and {Accurso}, G. and {Boissier}, J. and {Boone}, F. and {Bouch{\'e}}, N. and {Bournaud}, F. and {Burkert}, A. and {Carollo}, M. and {Cooper}, M. and {Cox}, P. and {Feruglio}, C. and {Freundlich}, J. and {Herrera-Camus}, R. and {Juneau}, S. and {Lippa}, M. and {Naab}, T. and {Renzini}, A. and {Salome}, P. and {Sternberg}, A. and {Tadaki}, K. and {{\"U}bler}, H. and {Walter}, F. and {Weiner}, B. and {Weiss}, A.},
        title = "{PHIBSS: Unified Scaling Relations of Gas Depletion Time and Molecular Gas Fractions}",
      journal = {\apj},
         year = 2018,
        month = feb,
       volume = {853},
       number = {2},
          eid = {179},
        pages = {179},
          doi = {10.3847/1538-4357/aaa4b4},
archivePrefix = {arXiv},
       eprint = {1702.01140},
 primaryClass = {astro-ph.GA},
       adsurl = {https://ui.adsabs.harvard.edu/abs/2018ApJ...853..179T}
}

@ARTICLE{Whitmore1990,
       author = {{Whitmore}, Bradley C. and {Lucas}, Ray A. and {McElroy}, Douglas B. and {Steiman-Cameron}, Thomas Y. and {Sackett}, Penny D. and {Olling}, Rob P.},
        title = "{New Observations and a Photographic Atlas of Polar-Ring Galaxies}",
      journal = {\aj},
         year = 1990,
        month = nov,
       volume = {100},
        pages = {1489},
          doi = {10.1086/115614},
       adsurl = {https://ui.adsabs.harvard.edu/abs/1990AJ....100.1489W}
}

@ARTICLE{Moiseev2011,
       author = {{Moiseev}, Alexei V. and {Smirnova}, Ksenia I. and {Smirnova}, Aleksandrina A. and {Reshetnikov}, Vladimir P.},
        title = "{A new catalogue of polar-ring galaxies selected from the Sloan Digital Sky Survey}",
      journal = {\mnras},
         year = 2011,
        month = nov,
       volume = {418},
       number = {1},
        pages = {244-257},
          doi = {10.1111/j.1365-2966.2011.19479.x},
archivePrefix = {arXiv},
       eprint = {1107.1966},
 primaryClass = {astro-ph.CO},
       adsurl = {https://ui.adsabs.harvard.edu/abs/2011MNRAS.418..244M}
}

@ARTICLE{Mosenkov2024,
       author = {{Mosenkov}, A.~V. and {Bahr}, S.~K.~H. and {Reshetnikov}, V.~P. and {Shakespear}, Z. and {Smirnov}, D.~V.},
        title = "{The occurrence rate of galaxies with polar structures may be significantly underestimated}",
      journal = {\aap},
         year = 2024,
        month = jan,
       volume = {681},
          eid = {L15},
        pages = {L15},
          doi = {10.1051/0004-6361/202348494},
archivePrefix = {arXiv},
       eprint = {2311.03529},
 primaryClass = {astro-ph.GA},
       adsurl = {https://ui.adsabs.harvard.edu/abs/2024A&A...681L..15M}
}

@ARTICLE{Bournaud2003,
       author = {{Bournaud}, F. and {Combes}, F.},
        title = "{Formation of polar ring galaxies}",
      journal = {\aap},
         year = 2003,
        month = apr,
       volume = {401},
        pages = {817-833},
          doi = {10.1051/0004-6361:20030150},
archivePrefix = {arXiv},
       eprint = {astro-ph/0301391},
 primaryClass = {astro-ph},
       adsurl = {https://ui.adsabs.harvard.edu/abs/2003A&A...401..817B}
}

@ARTICLE{Reshetnikov1997,
       author = {{Reshetnikov}, V. and {Sotnikova}, N.},
        title = "{Global structure and formation of polar-ring galaxies.}",
      journal = {\aap},
         year = 1997,
        month = sep,
       volume = {325},
        pages = {933-942},
          doi = {10.48550/arXiv.astro-ph/9704047},
archivePrefix = {arXiv},
       eprint = {astro-ph/9704047},
 primaryClass = {astro-ph},
       adsurl = {https://ui.adsabs.harvard.edu/abs/1997A&A...325..933R}
}

@ARTICLE{Brook2008,
       author = {{Brook}, Chris B. and {Governato}, Fabio and {Quinn}, Thomas and {Wadsley}, James and {Brooks}, Alyson M. and {Willman}, Beth and {Stilp}, Adrienne and {Jonsson}, Patrik},
        title = "{The Formation of Polar Disk Galaxies}",
      journal = {\apj},
         year = 2008,
        month = dec,
       volume = {689},
       number = {2},
        pages = {678-686},
          doi = {10.1086/591489},
archivePrefix = {arXiv},
       eprint = {0802.1051},
 primaryClass = {astro-ph},
       adsurl = {https://ui.adsabs.harvard.edu/abs/2008ApJ...689..678B}
}

@ARTICLE{Whitney2020,
       author = {{Whitney}, A. and {Conselice}, C.~J. and {Duncan}, K. and {Spitler}, L.~R.},
        title = "{Surface Brightness Evolution of Galaxies in the CANDELS GOODS Fields up to z {\ensuremath{\sim}} 6: High-z Galaxies Are Unique or Remain Undetected}",
      journal = {\apj},
         year = 2020,
        month = nov,
       volume = {903},
       number = {1},
          eid = {14},
        pages = {14},
          doi = {10.3847/1538-4357/abb824},
archivePrefix = {arXiv},
       eprint = {2009.07295},
 primaryClass = {astro-ph.GA},
       adsurl = {https://ui.adsabs.harvard.edu/abs/2020ApJ...903...14W}
}

@ARTICLE{Euclid2025VIS,
       author = {{Euclid Collaboration} and {Cropper}, M.~S. and {Al-Bahlawan}, A. and {Amiaux}, J. and {Awan}, S. and {Azzollini}, R. and {Benson}, K. and {Berthe}, M. and {Boucher}, J. and {Bozzo}, E. and {Brockley-Blatt}, C. and {Candini}, G.~P. and {Cara}, C. and {Chaudery}, R.~A. and {Cole}, R.~E. and {Danto}, P. and {Denniston}, J. and {Di Giorgio}, A.~M. and {Dryer}, B. and {Dubois}, J.-P. and {Endicott}, J. and {Farina}, M. and {Galli}, E. and {Genolet}, L. and {Gow}, J.~P.~D. and {Guttridge}, P. and {Hailey}, M. and {Hall}, D. and {Harper}, C. and {Hoekstra}, H. and {Holland}, A.~D. and {Horeau}, B. and {Hu}, D. and {James}, R.~E. and {Khalil}, A. and {King}, R. and {Kitching}, T. and {Kohley}, R. and {Larcheveque}, C. and {Lawrenson}, A. and {Liebing}, P. and {Liu}, S.~J. and {Martignac}, J. and {Massey}, R. and {McCracken}, H.~J. and {Miller}, L. and {Murray}, N. and {Nakajima}, R. and {Niemi}, S.-M. and {Nightingale}, J.~W. and {Paltani}, S. and {Pendem}, A. and {Philippon}, A. and {Plana}, C. and {Pool}, P. and {Pottinger}, S. and {Racca}, G.~D. and {Rhodes}, J. and {Rousseau}, A. and {Ruane}, K. and {Salatti}, M. and {Salvignol}, J.-C. and {Sciortino}, A. and {Short}, A. and {Skottfelt}, J. and {Smit}, S.~J.~A. and {Swindells}, I. and {Szafraniec}, M. and {Thomas}, P.~D. and {Thomas}, W. and {Tommasi}, E. and {Tosti}, S. and {Visticot}, F. and {Walton}, D.~M. and {Willis}, G. and {Winter}, B. and {Aghanim}, N. and {Altieri}, B. and {Amara}, A. and {Andreon}, S. and {Auricchio}, N. and {Aussel}, H. and {Baccigalupi}, C. and {Baldi}, M. and {Balestra}, A. and {Bardelli}, S. and {Basset}, A. and {Bender}, R. and {Bernardeau}, F. and {Bodendorf}, C. and {Boenke}, T. and {Bonino}, D. and {Branchini}, E. and {Brescia}, M. and {Brinchmann}, J. and {Camera}, S. and {Capobianco}, V. and {Carbone}, C. and {Cardone}, V.~F. and {Carretero}, J. and {Casas}, R. and {Casas}, S. and {Castander}, F.~J. and {Castellano}, M. and {Castignani}, G. and {Cavuoti}, S. and {Cimatti}, A. and {Colodro-Conde}, C. and {Congedo}, G. and {Conselice}, C.~J. and {Conversi}, L. and {Copin}, Y. and {Courbin}, F. and {Courtois}, H.~M. and {Crocce}, M. and {Cuby}, J.-G. and {Cuillandre}, J.-C. and {Da Silva}, A. and {Degaudenzi}, H. and {De Lucia}, G. and {Dinis}, J. and {Dolding}, C. and {Douspis}, M. and {Duncan}, C.~A.~J. and {Dupac}, X. and {Dusini}, S. and {Ealet}, A. and {Fabricius}, M. and {Farrens}, S. and {Ferriol}, S. and {Fosalba}, P. and {Fotopoulou}, S. and {Frailis}, M. and {Franceschi}, E. and {Franzetti}, P. and {Frugier}, P.-A. and {Fumana}, M. and {Galeotta}, S. and {Garilli}, B. and {George}, K. and {Gillard}, W. and {Gillis}, B. and {Giocoli}, C. and {G{\'o}mez-Alvarez}, P. and {Granett}, B.~R. and {Grazian}, A. and {Grupp}, F. and {Guzzo}, L. and {Haugan}, S.~V.~H. and {Herent}, O. and {Hoar}, J. and {Holliman}, M.~S. and {Holmes}, W. and {Hook}, I. and {Hormuth}, F. and {Hornstrup}, A. and {Hudelot}, P. and {Ili{\'c}}, S. and {Jahnke}, K. and {Jhabvala}, M. and {Joachimi}, B. and {Keih{\"a}nen}, E. and {Kermiche}, S. and {Kiessling}, A. and {Kilbinger}, M. and {Kubik}, B. and {Kuijken}, K. and {K{\"u}mmel}, M. and {Kunz}, M. and {Kurki-Suonio}, H. and {Lahav}, O. and {Laureijs}, R. and {Ligori}, S. and {Lilje}, P.~B. and {Lindholm}, V. and {Lloro}, I. and {Lorenzo Alvarez}, J. and {Mainetti}, G. and {Maino}, D. and {Maiorano}, E. and {Mansutti}, O. and {Marcin}, S. and {Marggraf}, O. and {Markovic}, K. and {Martinelli}, M. and {Martinet}, N. and {Marulli}, F. and {Masters}, D.~C. and {Maurogordato}, S. and {Medinaceli}, E. and {Mei}, S. and {Melchior}, M. and {Mellier}, Y. and {Meneghetti}, M. and {Merlin}, E. and {Meylan}, G. and {Mohr}, J.~J. and {Moresco}, M. and {Moscardini}, L. and {Neissner}, C.},
        title = "{Euclid: II. The VIS instrument}",
      journal = {\aap},
         year = 2025,
        month = may,
       volume = {697},
          eid = {A2},
        pages = {A2},
          doi = {10.1051/0004-6361/202450996},
archivePrefix = {arXiv},
       eprint = {2405.13492},
 primaryClass = {astro-ph.IM},
       adsurl = {https://ui.adsabs.harvard.edu/abs/2025A&A...697A...2E}
}

@ARTICLE{Reshetnikov2007,
       author = {{Reshetnikov}, V.~P. and {Dettmar}, R.-J.},
        title = "{HUDF 1619{\textemdash}A candidate for polar-ring galaxies in the Hubble Ultra Deep Field}",
      journal = {Astronomy Letters},
         year = 2007,
        month = apr,
       volume = {33},
       number = {4},
        pages = {222-226},
          doi = {10.1134/S1063773707040020},
archivePrefix = {arXiv},
       eprint = {astro-ph/0703784},
 primaryClass = {astro-ph},
       adsurl = {https://ui.adsabs.harvard.edu/abs/2007AstL...33..222R}
}

@ARTICLE{Euclid2025FWHM,
       author = {{Euclid Collaboration} and {Quilley}, L. and {Damjanov}, I. and {de Lapparent}, V. and {Paulino-Afonso}, A. and {Dom{\'\i}nguez S{\'a}nchez}, H. and {Ferr{\'e}-Mateu}, A. and {Huertas-Company}, M. and {K{\"u}mmel}, M. and {Delley}, D. and {Spiniello}, C. and {Baes}, M. and {Wang}, L. and {Kuchner}, U. and {Tarsitano}, F. and {Ragusa}, R. and {Siudek}, M. and {Tortora}, C. and {Aghanim}, N. and {Altieri}, B. and {Amara}, A. and {Andreon}, S. and {Auricchio}, N. and {Aussel}, H. and {Baccigalupi}, C. and {Baldi}, M. and {Balestra}, A. and {Bardelli}, S. and {Battaglia}, P. and {Bender}, R. and {Biviano}, A. and {Bonchi}, A. and {Bonino}, D. and {Branchini}, E. and {Brescia}, M. and {Brinchmann}, J. and {Camera}, S. and {Ca{\~n}as-Herrera}, G. and {Capobianco}, V. and {Carbone}, C. and {Carretero}, J. and {Casas}, S. and {Castander}, F.~J. and {Castellano}, M. and {Castignani}, G. and {Cavuoti}, S. and {Chambers}, K.~C. and {Cimatti}, A. and {Colodro-Conde}, C. and {Congedo}, G. and {Conselice}, C.~J. and {Conversi}, L. and {Copin}, Y. and {Costille}, A. and {Courbin}, F. and {Courtois}, H.~M. and {Cropper}, M. and {Da Silva}, A. and {Degaudenzi}, H. and {De Lucia}, G. and {Di Giorgio}, A.~M. and {Dolding}, C. and {Dole}, H. and {Duncan}, C.~A.~J. and {Dupac}, X. and {Dusini}, S. and {Escoffier}, S. and {Fabricius}, M. and {Farina}, M. and {Farinelli}, R. and {Faustini}, F. and {Ferriol}, S. and {Fosalba}, P. and {Fotopoulou}, S. and {Frailis}, M. and {Franceschi}, E. and {Galeotta}, S. and {George}, K. and {Gillis}, B. and {Giocoli}, C. and {Gracia-Carpio}, J. and {Granett}, B.~R. and {Grazian}, A. and {Grupp}, F. and {Haugan}, S.~V.~H. and {Hoekstra}, H. and {Holmes}, W. and {Hook}, I.~M. and {Hormuth}, F. and {Hornstrup}, A. and {Hudelot}, P. and {Jahnke}, K. and {Jhabvala}, M. and {Joachimi}, B. and {Keih{\"a}nen}, E. and {Kermiche}, S. and {Kiessling}, A. and {Kubik}, B. and {Kunz}, M. and {Kurki-Suonio}, H. and {Le Boulc'h}, Q. and {Le Brun}, A.~M.~C. and {Le Mignant}, D. and {Liebing}, P. and {Ligori}, S. and {Lilje}, P.~B. and {Lindholm}, V. and {Lloro}, I. and {Mainetti}, G. and {Maino}, D. and {Maiorano}, E. and {Mansutti}, O. and {Marcin}, S. and {Marggraf}, O. and {Martinelli}, M. and {Martinet}, N. and {Marulli}, F. and {Massey}, R. and {Maurogordato}, S. and {McCracken}, H.~J. and {Medinaceli}, E. and {Mei}, S. and {Melchior}, M. and {Mellier}, Y. and {Meneghetti}, M. and {Merlin}, E. and {Meylan}, G. and {Mora}, A. and {Moresco}, M. and {Moscardini}, L. and {Nakajima}, R. and {Neissner}, C. and {Niemi}, S. -M. and {Nightingale}, J.~W. and {Padilla}, C. and {Paltani}, S. and {Pasian}, F. and {Pedersen}, K. and {Percival}, W.~J. and {Pettorino}, V. and {Pires}, S. and {Polenta}, G. and {Poncet}, M. and {Popa}, L.~A. and {Pozzetti}, L. and {Raison}, F. and {Rebolo}, R. and {Renzi}, A. and {Rhodes}, J. and {Riccio}, G. and {Romelli}, E. and {Roncarelli}, M. and {Saglia}, R. and {Sakr}, Z. and {S{\'a}nchez}, A.~G. and {Sapone}, D. and {Sartoris}, B. and {Schewtschenko}, J.~A. and {Schneider}, P. and {Scodeggio}, M. and {Secroun}, A. and {Seidel}, G. and {Seiffert}, M. and {Serrano}, S. and {Simon}, P. and {Sirignano}, C. and {Sirri}, G. and {Stanco}, L. and {Steinwagner}, J. and {Tallada-Cresp{\'\i}}, P. and {Taylor}, A.~N. and {Teplitz}, H.~I. and {Tereno}, I. and {Tessore}, N. and {Toft}, S. and {Toledo-Moreo}, R. and {Torradeflot}, F. and {Tutusaus}, I. and {Valenziano}, L. and {Valiviita}, J. and {Vassallo}, T. and {Verdoes Kleijn}, G. and {Veropalumbo}, A. and {Wang}, Y. and {Weller}, J. and {Zacchei}, A. and {Zamorani}, G. and {Zerbi}, F.~M. and {Zinchenko}, I.~A. and {Zucca}, E. and {Allevato}, V. and {Ballardini}, M. and {Bolzonella}, M. and {Bozzo}, E. and {Burigana}, C. and {Cabanac}, R. and {Cappi}, A. and {Di Ferdinando}, D. and {Escartin Vigo}, J.~A. and {Gabarra}, L.},
        title = "{Euclid Quick Data Release (Q1). Exploring galaxy morphology across cosmic time through Sersic fits}",
      journal = {arXiv e-prints},
         year = 2025,
        month = mar,
          eid = {arXiv:2503.15309},
        pages = {arXiv:2503.15309},
          doi = {10.48550/arXiv.2503.15309},
archivePrefix = {arXiv},
       eprint = {2503.15309},
 primaryClass = {astro-ph.GA},
       adsurl = {https://ui.adsabs.harvard.edu/abs/2025arXiv250315309E}
}

@ARTICLE{Reshetnikov2015,
       author = {{Reshetnikov}, V.~P. and {Savchenko}, S.~S. and {Mosenkov}, A.~V. and {Sotnikova}, N. Ya. and {Bizyaev}, D.~V.},
        title = "{Polar-bulge galaxies}",
      journal = {Astronomy Letters},
         year = 2015,
        month = dec,
       volume = {41},
       number = {12},
        pages = {748-756},
          doi = {10.1134/S1063773715120117},
archivePrefix = {arXiv},
       eprint = {1511.05416},
 primaryClass = {astro-ph.GA},
       adsurl = {https://ui.adsabs.harvard.edu/abs/2015AstL...41..748R}
}

@ARTICLE{Corsini2012,
       author = {{Corsini}, E.~M. and {M{\'e}ndez-Abreu}, J. and {Pastorello}, N. and {Dalla Bont{\`a}}, E. and {Morelli}, L. and {Beifiori}, A. and {Pizzella}, A. and {Bertola}, F.},
        title = "{Polar bulges and polar nuclear discs: the case of NGC 4698}",
      journal = {MNRAS},
         year = 2012,
        month = jun,
       volume = {423},
       number = {1},
        pages = {L79-L83},
          doi = {10.1111/j.1745-3933.2012.01261.x},
archivePrefix = {arXiv},
       eprint = {1204.2265},
 primaryClass = {astro-ph.CO},
       adsurl = {https://ui.adsabs.harvard.edu/abs/2012MNRAS.423L..79C}
}

@ARTICLE{Mosenkov2020,
       author = {{Mosenkov}, Aleksandr and {Rich}, R. Michael and {Koch}, Andreas and {Brosch}, Noah and {Thilker}, David and {Rom{\'a}n}, Javier and {M{\"u}ller}, Oliver and {Smirnov}, Anton and {Usachev}, Pavel},
        title = "{The haloes and environments of nearby galaxies (HERON) - II. The outer structure of edge-on galaxies}",
      journal = {MNRAS},
         year = 2020,
        month = may,
       volume = {494},
       number = {2},
        pages = {1751-1770},
          doi = {10.1093/mnras/staa678},
archivePrefix = {arXiv},
       eprint = {2003.03392},
 primaryClass = {astro-ph.GA},
       adsurl = {https://ui.adsabs.harvard.edu/abs/2020MNRAS.494.1751M}
}

@ARTICLE{Mosenkov2022,
       author = {{Mosenkov}, Aleksandr V. and {Reshetnikov}, Vladimir P. and {Skryabina}, Maria N. and {Shakespear}, Zacory},
        title = "{Unveiling the Nature of Polar-ring Galaxies from Deep Imaging}",
      journal = {Research in Astronomy and Astrophysics},
         year = 2022,
        month = nov,
       volume = {22},
       number = {11},
          eid = {115003},
        pages = {115003},
          doi = {10.1088/1674-4527/ac8d82},
archivePrefix = {arXiv},
       eprint = {2208.12943},
 primaryClass = {astro-ph.GA},
       adsurl = {https://ui.adsabs.harvard.edu/abs/2022RAA....22k5003M}
}

@ARTICLE{Ordenes2016,
       author = {{Ordenes-Brice{\~n}o}, Yasna and {Georgiev}, Iskren Y. and {Puzia}, Thomas H. and {Goudfrooij}, Paul and {Arnaboldi}, Magda},
        title = "{Compact stellar systems in the polar ring galaxies NGC 4650A and NGC 3808B: Clues to polar disk formation}",
      journal = {\aap},
         year = 2016,
        month = jan,
       volume = {585},
          eid = {A156},
        pages = {A156},
          doi = {10.1051/0004-6361/201527025},
archivePrefix = {arXiv},
       eprint = {1511.05166},
 primaryClass = {astro-ph.GA},
       adsurl = {https://ui.adsabs.harvard.edu/abs/2016A&A...585A.156O}
}

@ARTICLE{Quiroga2020,
       author = {{Quiroga}, Luis F. and {Mu{\~n}oz-Cuartas}, J.~C. and {Rodrigues}, I. and {Libeskind}, Noam I.},
        title = "{Simulating the galactic system in interaction AM 2229-735 and the formation of its polar structure}",
      journal = {\mnras},
         year = 2020,
        month = jan,
       volume = {491},
       number = {2},
        pages = {1887-1900},
          doi = {10.1093/mnras/stz2943},
archivePrefix = {arXiv},
       eprint = {1911.03595},
 primaryClass = {astro-ph.GA},
       adsurl = {https://ui.adsabs.harvard.edu/abs/2020MNRAS.491.1887Q}
}

@ARTICLE{Martinez2010,
       author = {{Mart{\'\i}nez-Delgado}, David and {Gabany}, R. Jay and {Crawford}, Ken and {Zibetti}, Stefano and {Majewski}, Steven R. and {Rix}, Hans-Walter and {Fliri}, J{\"u}rgen and {Carballo-Bello}, Julio A. and {Bardalez-Gagliuffi}, Daniella C. and {Pe{\~n}arrubia}, Jorge and {Chonis}, Taylor S. and {Madore}, Barry and {Trujillo}, Ignacio and {Schirmer}, Mischa and {McDavid}, David A.},
        title = "{Stellar Tidal Streams in Spiral Galaxies of the Local Volume: A Pilot Survey with Modest Aperture Telescopes}",
      journal = {\aj},
         year = 2010,
        month = oct,
       volume = {140},
       number = {4},
        pages = {962-967},
          doi = {10.1088/0004-6256/140/4/962},
archivePrefix = {arXiv},
       eprint = {1003.4860},
 primaryClass = {astro-ph.CO},
       adsurl = {https://ui.adsabs.harvard.edu/abs/2010AJ....140..962M}
}

@ARTICLE{Khalid2024,
       author = {{Khalid}, A. and {Brough}, S. and {Martin}, G. and {Kimmig}, L.~C. and {Lagos}, C.~D.~P. and {Remus}, R.-S. and {Martinez-Lombilla}, C.},
        title = "{Characterizing tidal features around galaxies in cosmological simulations}",
      journal = {\mnras},
         year = 2024,
        month = jun,
       volume = {530},
       number = {4},
        pages = {4422-4445},
          doi = {10.1093/mnras/stae1064},
archivePrefix = {arXiv},
       eprint = {2404.12436},
 primaryClass = {astro-ph.GA},
       adsurl = {https://ui.adsabs.harvard.edu/abs/2024MNRAS.530.4422K}
}

@ARTICLE{Danovich2015,
       author = {{Danovich}, Mark and {Dekel}, Avishai and {Hahn}, Oliver and {Ceverino}, Daniel and {Primack}, Joel},
        title = "{Four phases of angular-momentum buildup in high-z galaxies: from cosmic-web streams through an extended ring to disc and bulge}",
      journal = {\mnras},
         year = 2015,
        month = may,
       volume = {449},
       number = {2},
        pages = {2087-2111},
          doi = {10.1093/mnras/stv270},
archivePrefix = {arXiv},
       eprint = {1407.7129},
 primaryClass = {astro-ph.GA},
       adsurl = {https://ui.adsabs.harvard.edu/abs/2015MNRAS.449.2087D}
}

@ARTICLE{Smirnova2013,
       author = {{Smirnova}, K.~I. and {Moiseev}, A.~V.},
        title = "{Are polar rings indeed polar?}",
      journal = {Astrophysical Bulletin},
         year = 2013,
        month = oct,
       volume = {68},
       number = {4},
        pages = {371-380},
          doi = {10.1134/S1990341313040019},
archivePrefix = {arXiv},
       eprint = {1311.4138},
 primaryClass = {astro-ph.GA},
       adsurl = {https://ui.adsabs.harvard.edu/abs/2013AstBu..68..371S}
}

@ARTICLE{Savchenko2017,
       author = {{Savchenko}, S.~S. and {Reshetnikov}, V.~P.},
        title = "{Spatial environment of polar-ring galaxies from the SDSS}",
      journal = {Astronomy Letters},
         year = 2017,
        month = mar,
       volume = {43},
       number = {3},
        pages = {146-151},
          doi = {10.1134/S1063773717020050},
archivePrefix = {arXiv},
       eprint = {1701.07978},
 primaryClass = {astro-ph.GA},
       adsurl = {https://ui.adsabs.harvard.edu/abs/2017AstL...43..146S}
}

@ARTICLE{Finkelman2012,
       author = {{Finkelman}, Ido and {Funes}, Jos{\'e} G. and {Brosch}, Noah},
        title = "{Polar ring galaxies in the Galaxy Zoo}",
      journal = {\mnras},
         year = 2012,
        month = may,
       volume = {422},
       number = {3},
        pages = {2386-2398},
          doi = {10.1111/j.1365-2966.2012.20790.x},
archivePrefix = {arXiv},
       eprint = {1202.5033},
 primaryClass = {astro-ph.CO},
       adsurl = {https://ui.adsabs.harvard.edu/abs/2012MNRAS.422.2386F}
}

@ARTICLE{Nelson2024,
       author = {{Nelson}, Erica and {Brammer}, Gabriel and {Gim{\'e}nez-Arteaga}, Clara and {Oesch}, Pascal A. and {Naidu}, Rohan P. and {{\"U}bler}, Hannah and {Matharu}, Jasleen and {Shapley}, Alice E. and {Whitaker}, Katherine E. and {Wisnioski}, Emily and {F{\"o}rster Schreiber}, Natascha M. and {Smit}, Renske and {van Dokkum}, Pieter and {Chisholm}, John and {Endsley}, Ryan and {Hartley}, Abigail I. and {Gibson}, Justus and {Giovinazzo}, Emma and {Illingworth}, Garth and {Labbe}, Ivo and {Maseda}, Michael V. and {Matthee}, Jorryt and {Covelo Paz}, Alba and {Price}, Sedona H. and {Reddy}, Naveen A. and {Shivaei}, Irene and {Weibel}, Andrea and {Wuyts}, Stijn and {Xiao}, Mengyuan and {Alberts}, Stacey and {Baker}, William M. and {Bunker}, Andrew J. and {Cameron}, Alex J. and {Charlot}, Stephane and {Eisenstein}, Daniel J. and {de Graaff}, Anna and {Ji}, Zhiyuan and {Johnson}, Benjamin D. and {Jones}, Gareth C. and {Maiolino}, Roberto and {Robertson}, Brant and {Sandles}, Lester and {Suess}, Katherine A. and {Tacchella}, Sandro and {Williams}, Christina C. and {Witstok}, Joris},
        title = "{Ionized Gas Kinematics with FRESCO: An Extended, Massive, Rapidly Rotating Galaxy at z = 5.4}",
      journal = {\apjl},
         year = 2024,
        month = dec,
       volume = {976},
       number = {2},
          eid = {L27},
        pages = {L27},
          doi = {10.3847/2041-8213/ad7b17},
archivePrefix = {arXiv},
       eprint = {2310.06887},
 primaryClass = {astro-ph.GA},
       adsurl = {https://ui.adsabs.harvard.edu/abs/2024ApJ...976L..27N}
}

@misc{https://doi.org/10.26131/irsa602,
  doi = {10.26131/IRSA602},
  url = {https://catcopy.ipac.caltech.edu/dois/doi.php?id=10.26131/IRSA602},
  author = {{Euclid Consortium}},
  title = {Euclid Q1 MER Catalog},
  publisher = {IPAC},
  year = {2025}
}

\begin{appendix}

\section{Sky map}
\label{sec:sky map}

\begin{figure}[htbp]
    \centering
    \includegraphics[width=\linewidth, height=5.8cm]{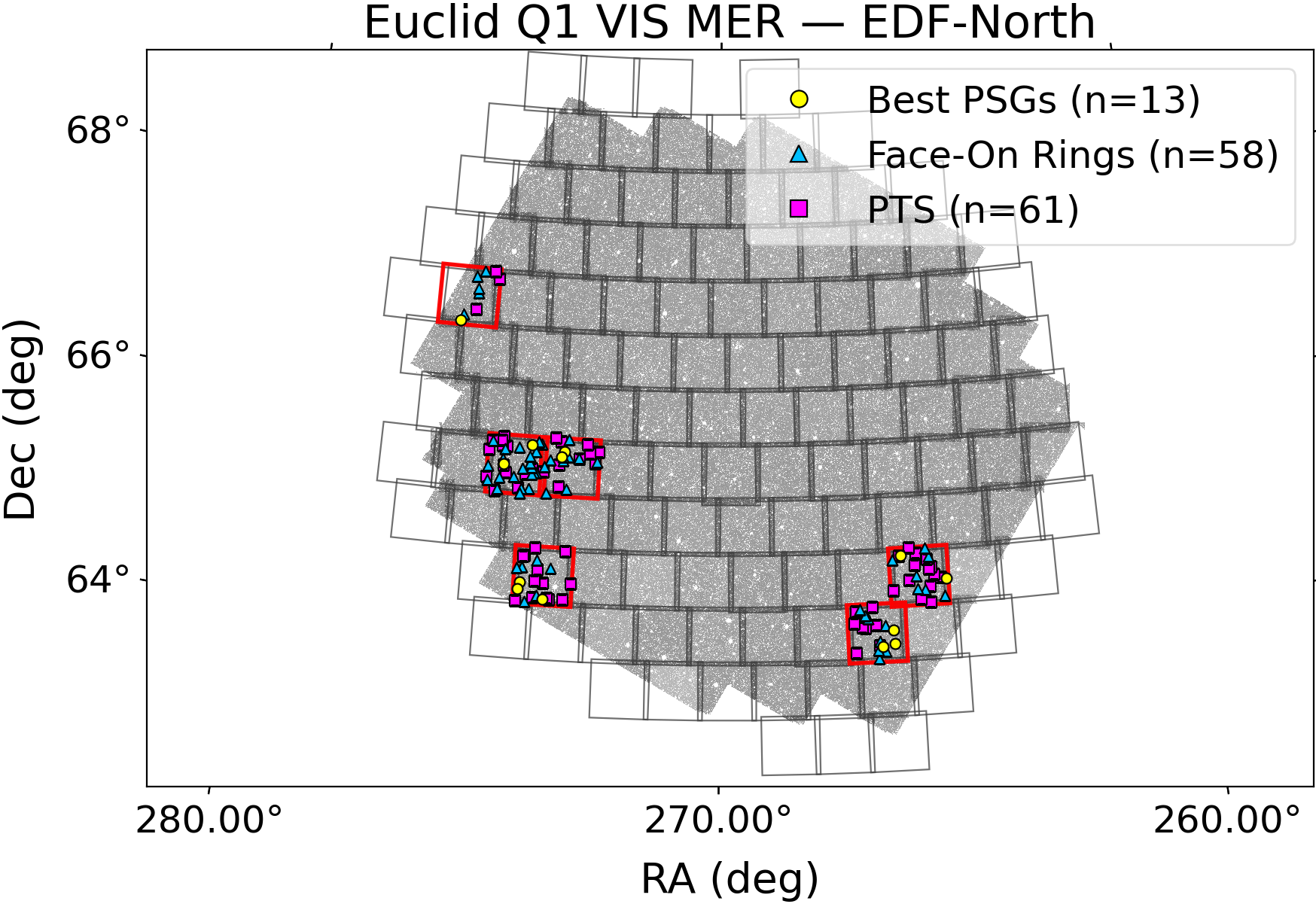}
    \includegraphics[width=\linewidth, height=5.8cm]{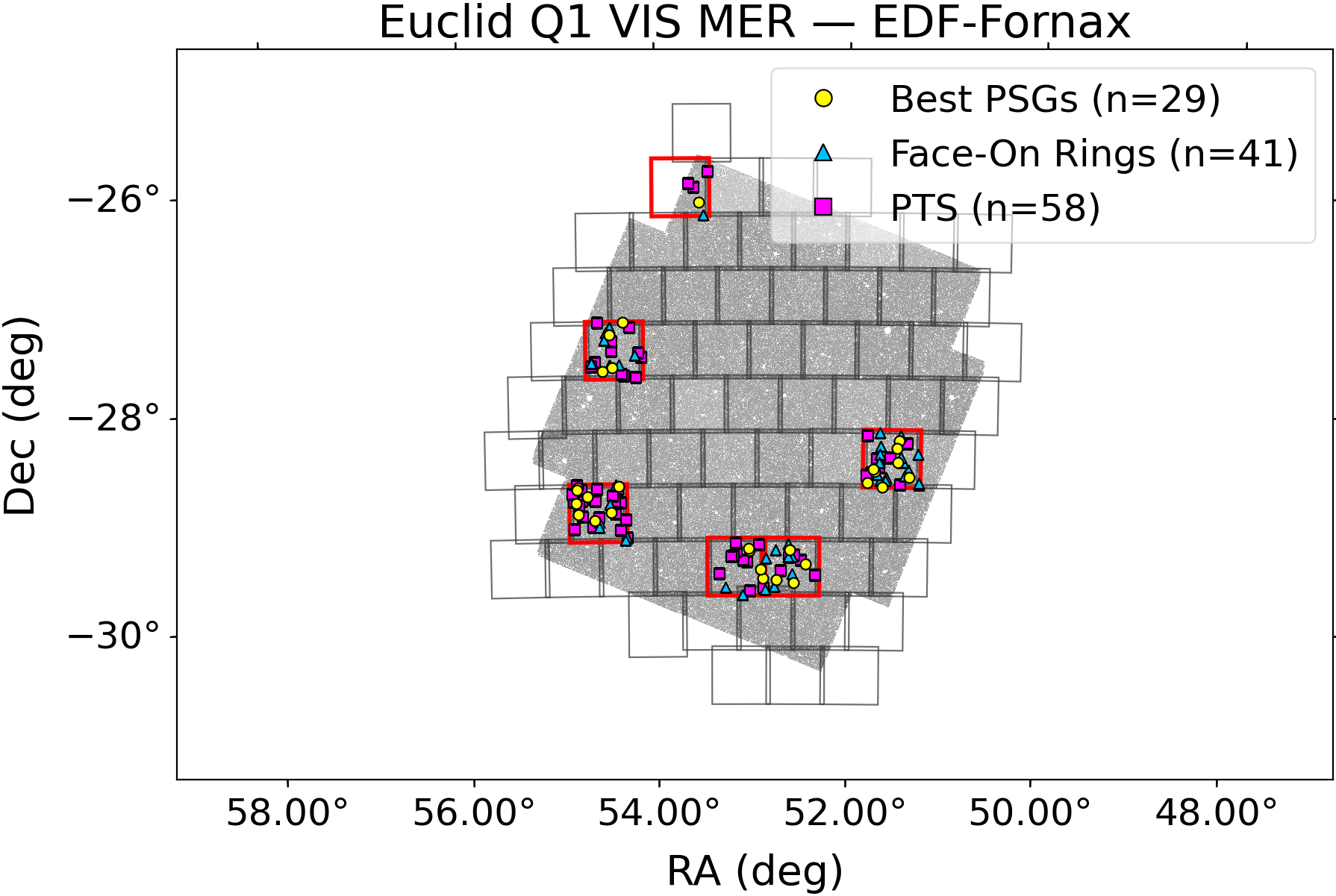}
    \includegraphics[width=\linewidth, height=5.8cm]{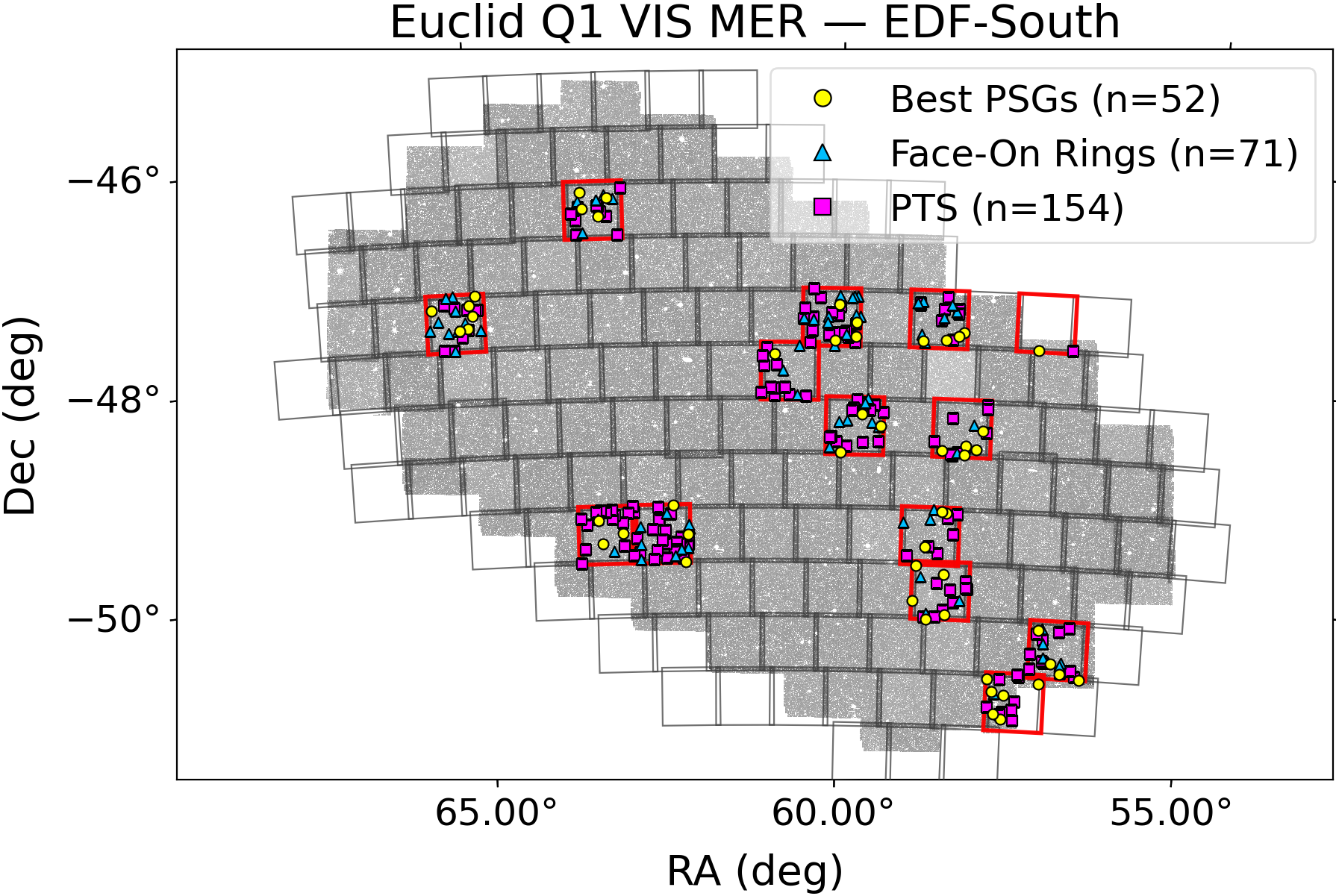}
    \caption{Sky maps of the three main Euclid deep fields: North, Fornax, and South. The 26 tiles used in this study are outlined in red. Every galaxy we identified is marked, with the markers and the number of each sample in each deep field expressed in the legends.}
    \label{fig:skymaps}
\end{figure}

The sky coverage of the three Euclid deep fields is shown in
Fig.~\ref{fig:skymaps}.

\section{Atlas Images}
\label{sec:Atlas}

The enhanced outer structure, enhanced outer structure + isophote map, and enhanced inner structure images in the atlas were all generated from the $I_{\mathrm{E}}$ filter data. The enhanced outer structure images used an SDSS-style RGB rendering, with the same $I_{\mathrm{E}}$ image supplied to the $g$, $r$, and $z$ channels. The channels were scaled by 6.0, 3.4, and 2.2, respectively, with a pedestal of 0.03. An asinh mapping with $Q=20$ was followed by an additional faint-feature asinh stretch with $Q=5$ and CLAHE enhancement using a clip limit of 3.3 and a $1\times1$ tile grid. The resulting intensity was inverted to produce the white background and orange/brown bright regions, emphasizing the low-surface-brightness outer structures.

The isophote maps were generated from the Euclid VIS $I_{\mathrm{E}}$ FITS cutouts by converting pixel intensities to surface brightness using a zeropoint of 24.6 and a pixel scale of $0.1''$ per pixel. Contours spanning $21$--$26~\mathrm{mag~arcsec^{-2}}$ were plotted at $0.5~\mathrm{mag~arcsec^{-2}}$ intervals, with Gaussian smoothing increasing linearly from $\sigma_{\min}=1.5$ pixels for the innermost isophote to $\sigma_{\max}=2.5$ pixels for the outermost. Isophotes that did not enclose any of the central $3\times3$ pixels were omitted to remove contours associated with unrelated sources. The isophote maps were then overlaid on the enhanced outer structure images.

\begin{figure}[htbp]
    \centering
    \includegraphics[width=0.48\linewidth]{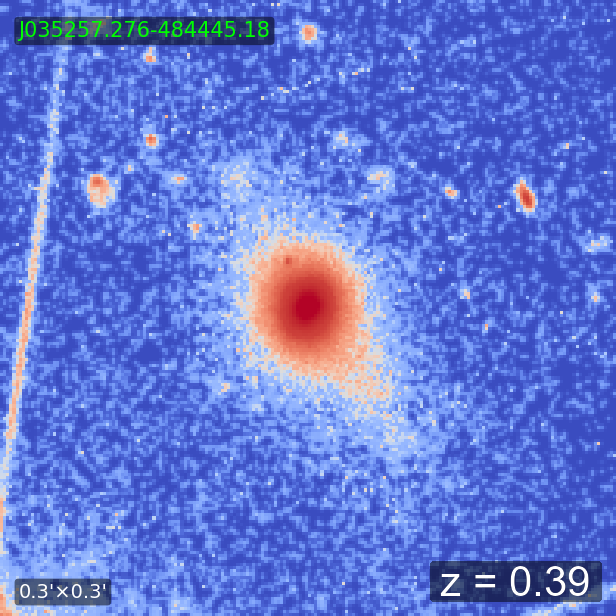}
    \includegraphics[width=0.48\linewidth]{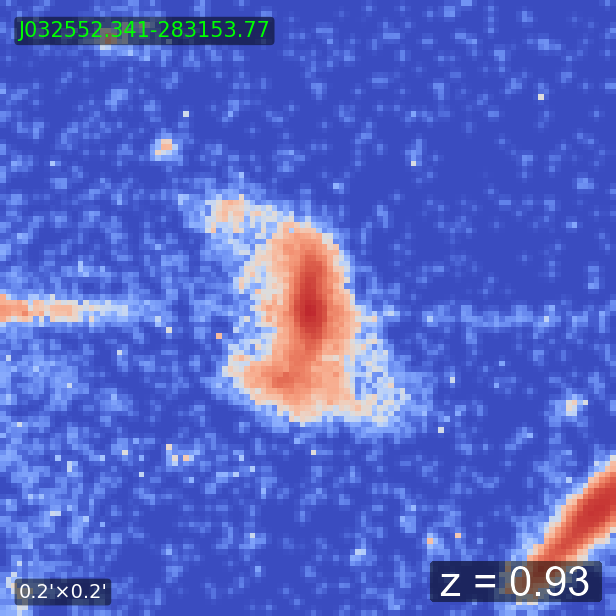}
    \\
    \includegraphics[width=0.48\linewidth]{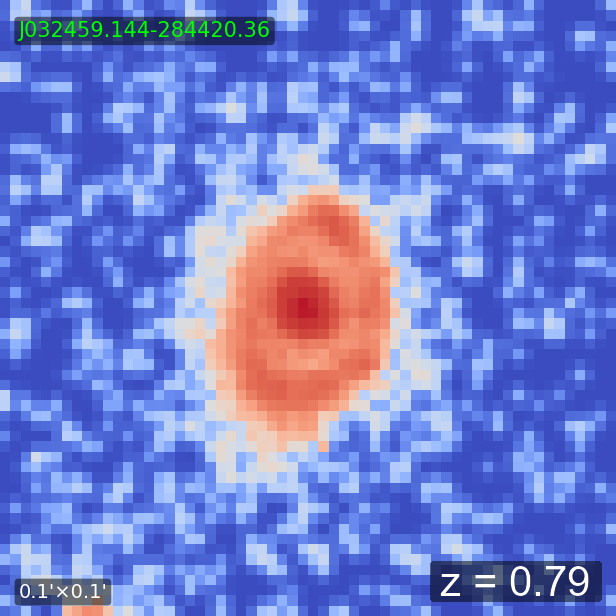}
    \includegraphics[width=0.48\linewidth]{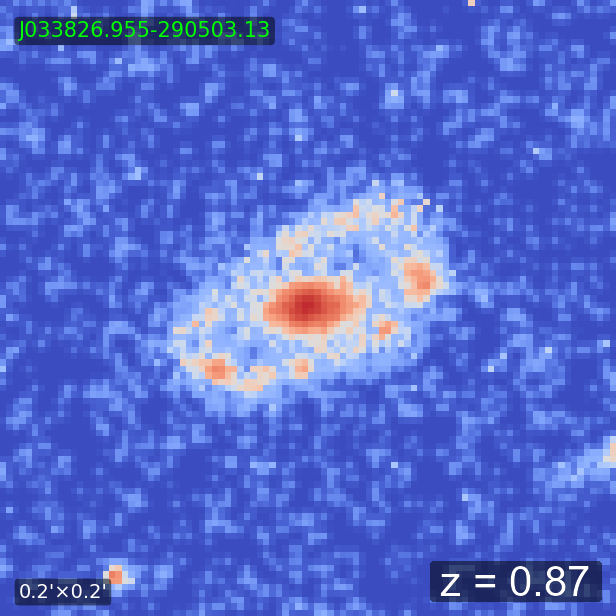}
    \\
    \includegraphics[width=0.48\linewidth]{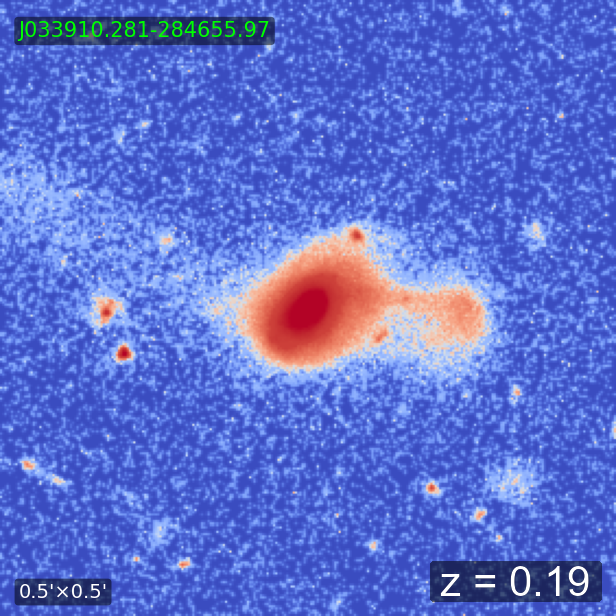}
    \includegraphics[width=0.48\linewidth]{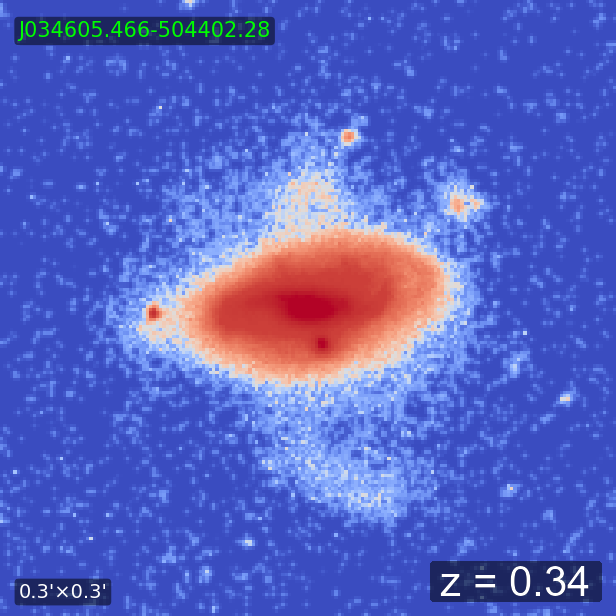}
    \caption{The same representative Euclid VIS cutouts from Fig.~\ref{fig:sample_gals}, shown in the enhanced-inner-structure style.}
    \label{fig:sample_inner_structure}
\end{figure}

Following the enhancement method described in Appendix A of \citet{Bahr2026}, the enhanced inner structure images used an asinh stretch for regions brighter than $\mu_r = 25.0~\mathrm{mag~arcsec^{-2}}$ ($\mathrm{bias}=0.23$, $\mathrm{contrast}=1.03$, $\mathrm{brightness}=1.72$, $\sigma=0$) and a logarithmic stretch for fainter regions ($\mathrm{bias}=0.18$, $\mathrm{contrast}=1.08$, $\mathrm{brightness}=1.11$, $\sigma=0.6$). The two branches were smoothly combined and displayed with the \texttt{coolwarm} colour map, using a zeropoint of 24.6 and a pixel scale of $0.1''$ per pixel. Examples of these images are shown in Fig.~\ref{fig:sample_inner_structure}.

\end{appendix}

\end{document}